\documentclass[fleqn,usenatbib]{mnras}

\usepackage{newtxtext,newtxmath}
\usepackage{epsf,rotating}
\usepackage{color}

\usepackage{verbatimbox}
\usepackage{bm}
\usepackage{cleveref}
\usepackage{ae,aecompl}
\usepackage{float}

\usepackage{graphicx}	
\usepackage{amsmath}	
\usepackage{amssymb}	

\title[EoR history CNN estimator]{Constraining the reionization history using deep learning from 21cm tomography with the Square Kilometre Array}

\author[Mangena, Hassan \& Santos]{
Tumelo  Mangena,$^{1,2}$\thanks{E-mail: tumelokholofelo@gmail.com}
Sultan Hassan,$^{3,1}\thanks{Tombaugh Fellow}$
Mario G. Santos$^{1,2}$
\\
$^{1}$ Department of Physics and Astronomy, University of the Western Cape, Bellville, Cape Town, 7535, South Africa\\
$^{2}$ South African Radio Astronomy Observatory (SARAO), 2 Fir Street, Observatory, Cape Town, 7925, South Africa\\
$^{3}$ Department of Astronomy, New Mexico State University, Las Cruces, NM, 88003, USA
}

\date{Accepted XXX. Received YYY; in original form ZZZ}

\pubyear{2019}

\begin{document}
\label{firstpage}
\pagerange{\pageref{firstpage}--\pageref{lastpage}}
\maketitle

\begin{abstract}
Upcoming 21cm surveys with the SKA1-LOW telescope will enable imaging of the neutral hydrogen distribution on cosmological scales in the early Universe. These surveys are expected to generate huge imaging datasets that will encode more information than the power spectrum. This provides an alternative unique way to constrain the reionization history, which might break the degeneracy in the power spectral analysis. Using Convolutional Neural Networks (CNN), we create a fast estimator of the neutral fraction from the 21cm maps that are produced by our large semi-numerical simulation. Our estimator is able to efficiently recover the neutral fraction ($x_{\rm HI}$) at several redshifts with a high accuracy of 99\%  as quantified by the coefficient of  determination $R^{2}$. Adding the instrumental effects from the SKA design slightly increases the loss function, but nevertheless we are still able to recover the neutral fraction with a similar high accuracy of 98\%, which is only 1 per cent less. While a weak dependence on redshift is observed, the accuracy increases rapidly with decreasing neutral fraction. This is due to the fact that the instrumental noise increases towards high redshift where the Universe is highly neutral.  Our results show the promise of directly using 21cm-tomography to constrain the reionization history in a model independent way, complementing similar efforts, such as those of the optical depth measurements from the  Cosmic Microwave Background (CMB) observations by {\it Planck}.
\end{abstract}




\section{Introduction}
The Epoch of Reionization (EoR) marks a period in the early Universe during which the birth of the first luminous cosmic structures gradually reionized the Inter-galactic medium (IGM). The EoR contains enormous cosmological and astrophysical information that is important to understand galaxy evolution and formation \citet{loeb2001reionization}. 

However, the history of reionization remains quite unconstrained. Lyman-$\alpha$ forest observations at z$\sim$6 have placed upper limits on the IGM neutral fraction, which indicates that the universe was nearly ionized by these epochs, $x_{\rm HI} \leq 0.01$~\citep{fan2006observational,becker2015evidence}.  Using identified Ly$\alpha$ emitters sample from the KMOS Lens-Amplified Spectroscopic Survey (KLASS),~\citet{mason2019} recently have been able to place a lower limit on the average IGM neutral hydrogen fraction of $> 0.76 \,(68\%),>0.46 \,(95\%)$ at z$\sim$8, indicating a rapid evolution at the end of reionization. Using the quasars damping wing analysis,~\citet{davis18} has constrained the IGM neutral fraction to be 0.6 at z=7.54 and 0.48 at z=7.09. A complementary similar analysis by~\citet{greig19} has suggested that the IGM neutral fraction is about 0.2 at z=7.5. On the other hand, the Cosmic Microwave Background (CMB) observations also provide constraints on the duration of reionization, through the integrated optical depth. The Wilkinson Microwave Anisotropy Prob~\citep[][WMAP]{hinshaw2013nine} CMB observations have previously measured an optical depth of 0.088, implying that reionization was complete by z$\sim$ 10. However, the much lower optical depth of 0.058 reported recently by {\it Planck} favors a sudden, short and late reionization by z$\sim$7-8. Their measurements suggest that the Universe was less than 10\% ionized by z$\sim$10. This shows the need for an additional probe that might break model degeneracy and provide an alternative and direct probe to the neutral fraction.

The 21cm hyperfine line carries a wealth of information, which is promising to unravel the IGM nature and hence the reionization history. Many radio interferometer experiments, such as the Low Frequency Array~\citep[LOFAR;][]{van2013lofar}, the Precision Array for Probing the Epoch of Reionization~\citep[PAPER;][]{parsons2010precision}, the Murchison Wide field Array~\citep[MWA;][]{bowman2013science}, the Giant Metrewave Radio Telescope~\citep[MWA;][]{paciga2011gmrt}, the Hydrogen Epoch of Reionization Array \citep[HERA;][]{deboer2017hydrogen} and Square Kilometer Array~\citep[SKA;][]{mellema2013reionization}, aim to detect reionization on cosmological scales through its 21 cm fluctuations. All these experiments motivate the preparation and development of theoretical methods and statistical tools to extract possible constraints on the reionization history.

In this light, many methods have been already developed to constrain the reionization history. These include: fitting to Ly-$\alpha$ and optical depth measurements using MCMC linked to a semi-numerical model~\citep[e.g.][]{greig2016global}, inferring the reionization history from parameters constraints against 21cm  mock power spectrum~\citep[e.g.][]{greig201521cmmc}, identifying HII bubbles~\citep[e.g.][]{Sambit2018} and recently using Convolutional neural networks (CNNs) to constrain the reionization duration~\citep{la2018machine}. While \cite{la2018machine} found that their designed CNN recovers the reionization duration within $\sim$ 10\%, their pipeline ignores the thermal noise contribution and implements a more simplified angular resolution treatment through applying a cut-off on the k$_{\perp}$ modes inferred from the experiment resolution. 

In this work, we design a CNN to extract the neutral fraction directly from 21cm-images at each redshift, thus producing a full reionization history, which is one of the first quantities we would like to measure from 21cm experiments. This approach relies directly on imaging and the relation between the 21cm signal and ionised patches, without requiring to go through the derivation of the ionization fraction from power spectrum statistics, being therefore more robust to model assumptions. The images are produced by our semi-numerical model {\sc SimFast21}\footnote{https://github.com/mariogrs/Simfast21} \citep{santos2010fast}. We implement a more physically motivated and realistic 21cm noise to our 21-images, following the recipe presented in \cite{hassan2018identifying}, that accounts for the experiment thermal noise, the angular resolution using the detailed baseline distribution as well as the effect of foregrounds. We focus our analysis on phase one of the SKA (the low-frequency part, SKA1-LOW), given its great imaging capabilities, although our approach can easily be applied to other 21cm arrays such as HERA and LOFAR. Our CNN is developed in python with the TensorFlow framework\footnote{https://www.tensorflow.org} ({\sc TensorFlow}), an open-source software library for numerical computation \citep{abadi2015tensorflow}.

This paper is organized as follows: we provide a summary of the 21cm brightness temperature semi-numerical simulations and the 21cm noise simulations in Section~\ref{sec:sim}. The designed CNN is introduced in Section~\ref{sec:cnn}. We present our results in Section~\ref{sec:results}, and conclude in Section~\ref{sec:conclusion}.

\section{Simulations}\label{sec:sim}

\subsection{\sc SimFast21}

We use the Instantaneous version of our semi-numerical model {\sc SimFast21} that has been developed in \cite{hassan2016simulating}. This model has been recently shown to be in a good agreement with predictions from our radiative transfer simulation~\citep[ARTIST;][]{molaro2019artist} in terms of the ionization morphology and 21cm power spectrum. We describe generally the simulation ingredients, and defer to \cite{santos2010fast} for the full details of the simulation algorithm, and to \cite{hassan2016simulating} for further updates on the Instantaneous model.

The initial step of the simulation is to generate the dark matter density from a Gaussian distribution using a Monte-Carlo approach in the linear regime. Next, it dynamically evolves the density field from the linear to non-linear regime by applying the \citet{zel1970gravitational} approximation. The dark matter halos are then generated using the excursion-set formalism (ESF). Ionized regions are identified using a similar form of the ESF that is based on a comparison between the  ionization rate $R_{\rm ion}$ and the recombination rate $R_{\rm rec}$ in spherical regions of decreasing sizes as specified by the ESF. Regions are flagged as ionized if:
\begin{equation}
\centering
f_{\rm esc}\,  R_{\rm ion} \geq R_{\rm rec},
\end{equation}
where $f_{\rm esc}$ is our assumed escape fraction. The $R_{\rm ion}$ parameterization is derived from a combination of the radiative transfer simulation \citep{finlator2015reionization}, and larger hydrodynamic simulation \citep{dave2013neutral}  that have both been shown to reproduce a wide range of observations, including low-redshift observations. The $R_{\rm ion}$ is parameterized as a function of halo mass $M_{\rm h}$ and redshift $z$ as follows:
\begin{equation}\label{eq:nion}
\frac{R_{\rm ion}}{M_{\rm h}} =  1.1 \times 10^{40} \,(1 + z)^{D_{\rm ion}} \left( \frac{M_{\rm h}}{9.51 \times 10^{7}}\right)^{C_{\rm ion}}\exp \left(  \frac{-9.51 \times 10^{7}}{M_{h}}\right)^{3.0},
\end{equation}
where the best fit parameters were found to be C$_{ion} = 0.41$ and D$_{ion} = 2.28$~\citep{hassan2016simulating}. Later, we will vary these parameters to generate our training dataset, which quantify the ionizing emissivity dependence on halo mass and redshift. Note that equation~\eqref{eq:nion} shows that $R_{\rm ion}$ scales as M$_{\rm h}^{1.41}$,  which is consistent with the SFR$-M_{\rm h}$ relation previously found by \cite{finlator2011galactic}. The $R_{\rm rec}$ is obtained from the radiative transfer simulation \citep{finlator2015reionization}, in order to account for the clumping effects below our cell size. The $R_{\rm rec}$ is parameterized as a function of over-density $\Delta$ and redshfit z as follows:
\begin{equation}
\frac{R_{\rm rec}}{V} =  9.85 \times 10^{-24} (1+ z)^{5.1}  \left[\frac{\left( \Delta/1.76 \right)^{0.82}}{1+ \left( \Delta/1.76 \right)^{0.82} } \right]^{4} \, , 
\end{equation}
where  V refers to the cell volume. We defer to \cite{hassan2016simulating} for the full details on the derivation of the $R_{\rm ion}$ and $R_{\rm rec}$ and their effects on several reionization observable.  The 21cm brightness temperature boxes are directly computed using the density and ionization fields assuming that the spin temperature is much higher than the CMB temperature, which is a valid assumption at lower redshifts, $z  <10$~\citep[e.g][]{santos2010fast} as considered in this work. 

\subsection{21cm Instrument Simulation}
The next step is to convert this simulation into a more realistic dataset by including instrumental effects (such as noise, resolution, foreground residuals). We call these new images, the "mock" maps. We partially follow the recipe developed in \citet{hassan2018identifying}, particularly for the foreground cleaning. We account directly for the resolution in frequency and angular scales following the most updated documentation for the SKA-LOW configuration~\citep{braun_ska}. According to the imaging sensitivity analysis performed by~\citet{braun_ska} in the frequency range of our interest (z=6-10, $\nu$=203-129 MHz), the line sensitivity is equal to 973 $\mu$Jy/beam for a  fractional bandwidth ($\Delta \nu / \nu$ = 10$^{-4}$) and integration time $t_{\rm int} =1$ hour. These values will be used directly in our noise pipeline to include more realistic instrumental effects following the most updated SKA-LOW design. Here we restrict our analysis to the SKA1-LOW design, given its promising imaging capabilities and defer a more detailed comparison between different experiments, such as with HERA and LOFAR, to future work. 

Using~{\sc SimFast21}, we run many large scale reionization simulations with a box size of 500 Mpc and 200$^{3}$ cells. This box size is sufficient to capture reionization on large scales~\citep{iliev14}. We then add a realistic SKA-like noise to these simulations using the following pipeline which consists of three parts:

\begin{table}\huge
 \scalebox{0.5}{\begin{tabular}{ l  c }\hline
    Array design & 512 stations \\ \hline
    Station diameter, D $[m]$ & 35 \\ \hline
    Total observation time t$_{\rm int}$ [h] & 3000  \\ \hline
    Line sensitivity [$\mu$Jy/beam] & 973 \\ \hline
    Map Resolution [arcmin] &  5  \\ \hline
    Frequency resolution [MHz] & 1 \\ \hline
    Redshift &  10, 9, 8 ,7, 6  \\ \hline
    Frequency [MHz] &  129 , 142, 158, 178, 203  \\ \hline
    Thermal noise [mK] & 1.8, 1.6, 1.4, 1.2, 1. \\ \hline
    Default wedge slope $m$, Equation~\eqref{eq:wedge_eq} & 0.27, 0.23, 0.19, 0.15\\ 
    \hline
\end{tabular}}
\caption{Summary of our assumed SKA design.}\label{array_tab}
\end{table}

\begin{itemize}
     
    \item Foreground cleaning: Fourier modes that lie outside the so-called reionization window (i.e. inside the foreground wedge) are foreground contaminated. The foreground wedge slope \textit{m} is approximately given by:
    \begin{equation}\label{eq:wedge_eq} 
    m = \frac{ D \, H_{0}\, E(z)\, \sin \theta}{c(1+z)},
    \end{equation}
    where $H_{0}$ is the Hubble parameter, $c$ is the speed of light, $E(z)\equiv\sqrt{\Omega_{m}(1+z)^{3} +\Omega_\Lambda}$, and $\theta$ is the field of view. We zero out all modes within the wedge, satisfying k$_{\parallel}$ < \textit{m} k$_{\perp}$. For the same experiment (same beam angle), the slope (\textit{m}) increases with redshift, which means more modes are removed at higher redshifts. We quote exact \textit{m} values for the SKA at our redshifts of interest in Table~\ref{array_tab}. This is the first part of the noise pipeline which is applied on all 21cm boxes to filter out the foreground wedge. 
    \item Resolution: our simulation has a frequency resolution, the slice width in $z$-direction, of about $\sim$ 0.15 MHz at z=8. We choose to apply the noise on maps with frequency resolution of 1 MHz, and hence we average 6 successive slices along $z$-axis to obtain a single mock map. The angular resolution of our simulation is about  $\sim$  1 arcmin at $z=8$, and we convert our maps on x-y plane from 200$\times$200 to 40$\times$40 pixels to obtain a resolution of 5 arcmin. This step is applied on the foreground filtered box from the previous step.
    
    \item Thermal noise: we use directly the line sensitivity reported by~\citet{braun_ska}, and then convert its value using the parameters assumed in our mock survey. We assume long integration time of 3,000 hours following the deep survey strategy outlined by~\citet{koop15}. For angular resolution of 5 arcmin, frequency resolution of 1 MHz, and integration time of 3,000 hours, the thermal noise is about $\sim$ 1 mK at z=6. All values at other redshifts are quoted in Table~\ref{array_tab}.
    We then generate an 40$\times$40 pixel map whose pixels' values are drawn from a Gaussian distribution with a zero mean and standard deviation that is set to the thermal noise value.
    We then finally add the thermal noise map to the resulting map from the previous step.

\end{itemize}

\begin{figure*}
    \centering
    \includegraphics[scale=1.2]{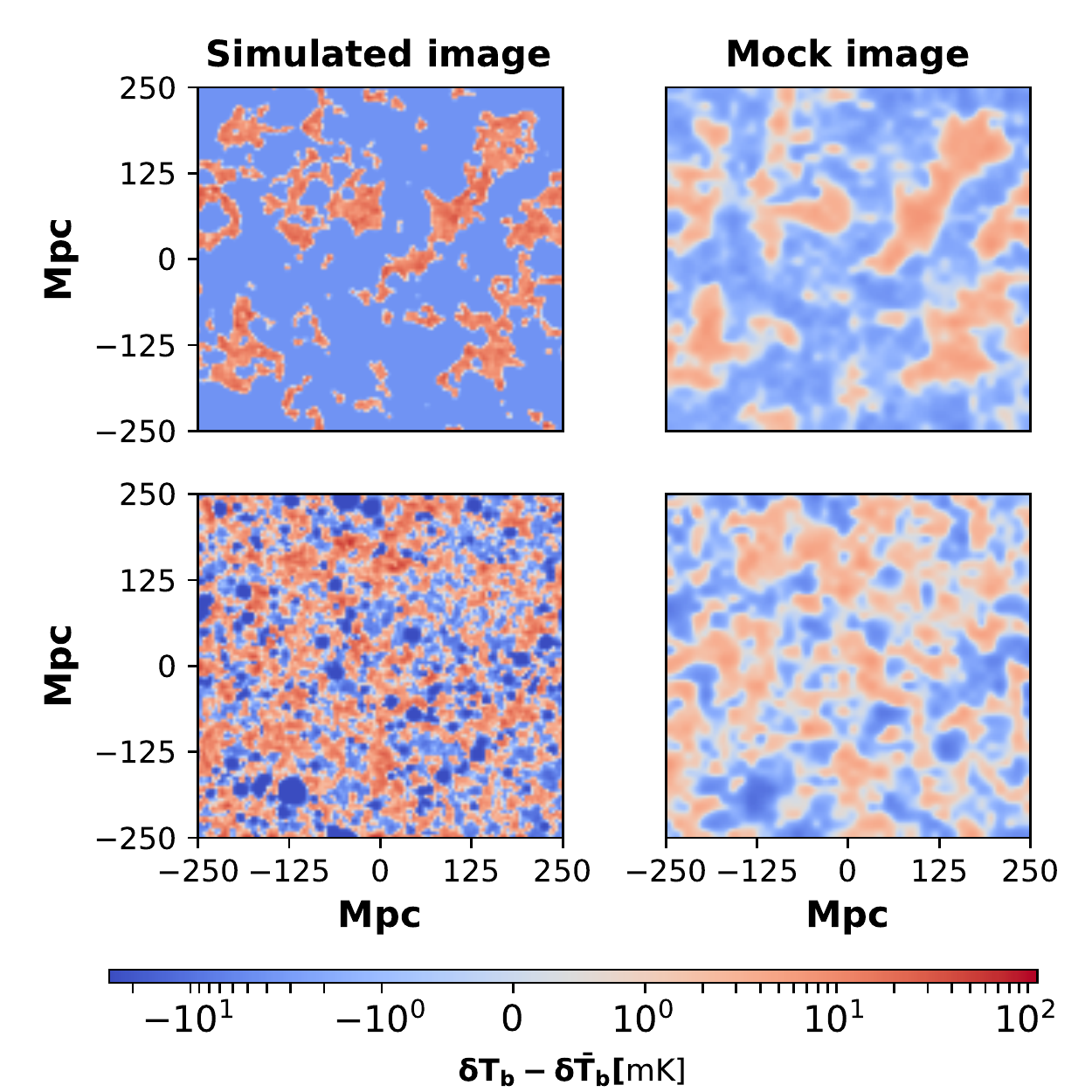}
    \caption{Example of two randomly selected 21cm images and their mock versions, using our assumed SKA design, at x$_{\rm HI} \sim$ 20\% (top) and 80\% (bottom). The mean value from each image is subtracted.}
    \label{fig:sample}
\end{figure*}
Figure~\ref{fig:sample} shows two randomly selected 21cm images from our training samples from a box size of 500 Mpc and N=200$^{3}$, at neutral fractions x$_{\rm HI} \sim$ 20\% (top), and 80\% (bottom) at z=6 and z=10, respectively. On the right, we show the corresponding 21cm mock images obtained using our 21cm noise pipeline as described above. The high angular resolution of our assumed SKA design, due to the high $uv$-coverage, allows resolving most of the original ionized bubbles on large and small scales. In highly ionized IGM (e.g. top panels), the noise dominates the ionized regions where the 21cm signal vanishes, but nevertheless, the signal features are still clearly present. 
In contrast, at highly neutral IGM (e.g. bottom panels), the ionized bubbles are very small, extending to the scales where the noise has the maximum effect. In this case, the original signal features are much harder to recognize, which might be more challenging for the neural network to derive exactly the true neutral fraction.

\section{Convolutional Neural network}\label{sec:cnn}

Convolutional Neural Networks (CNNs) have been very successful in reionization to  perform parameter regression \citep[e.g.][]{gillet2019deep,hassan2020}, classification \citep[e.g.][]{hassan2018identifying}, and emulating reionization simulation~\citep[e.g.][]{chardin19}. CNN is a subclass of neural networks - a system of interconnected artificial neurons that exchange messages between each other. It draws its power of best performance in dealing with images by taking advantage of the spatial structure of the inputs (for a comprehensive review see \citealt{rawat2017deep}). The basic structure of CNN consists of convolutional, pooling, and fully connected layers. Each layer is basically a linear combination of the components of the input (x) with weights and biases. Convolutional layers play the role of feature extraction from a 2D input. The role of the pooling layer is to up/down-sample the output of a convolutional layer, which basically reduces the spatial size and resolution of the features. The fully connected layers aim is the global features extraction from a 1D input.  The CNN attempts to find the best set of weights and biases that minimize a specific loss function, which is the distance between the true labels and the network predictions. The minimization occurs through an optimizer that computes the loss function gradients with respect to all network parameters (weights and biases). These gradients, in principle, indicate the directions at which the network parameters can be updated so that the loss is smaller, and hence the network predictions are closer to the true expected labels. Most optimizers are variants of the Stochastic Gradient Descent (SGD).  Our network uses Huber loss~\citep{hub64} as a loss function. Huber loss draws its power from combining advantages of two widely used loss functions, Mean Square Error (MSE) and Mean Absolute Error (MAE). It automatically turns into MAE when the error is large, and hence making it less sensitive to outliers. We use the Adaptive Moment Estimation~\citep[Adam,][]{kinba14} as an optimizer algorithm, which is a modified SGD form. Adam adds a momentum term to the classical SGD to accelerate convergence and adapts the learning rate automatically. These features ensures fast learning, while being less sensitive to the outliers. We present the summary of the hyper-parameters used in our network in Table~\ref{table:hyper}.

\begin{table}
    \caption{Summary of the hyper-parameters choice for the network} 
    \centering 
    \begin{tabular}{c | c} 
        \hline 
        Hyper-parameter & Value\\
        \hline
        Learning rate & start $= 10^{-2}$ \\
        Batch-size & 60 \\
        Number of Epochs & 300 \\
        Dropout rate & $20\% $ \\
        \hline 
    \end{tabular}
    \label{table:hyper} 
\end{table}

Our best performing network is composed of 3 convolutional blocks  and 3 fully connected layers. Each convolutional block contains 2 convolutional layers with 5$\times$5 kernel size and followed by 2D max-pooling layer. Batch normalization and the Rectified Linear Unit (ReLU) activation are applied after each layer, except the final output layer, to regularize the training process. A 20 \% dropout rate is applied after the first fully connected layers in order to prevent over-fitting. Summary of the architecture used in this work is presented in Table~\ref{table:arch}. Referring to Table~\ref{table:hyper}, we train the network for a total of 300 epochs with the start learning rate of $10^{-2}$.

\begin{table}
    \caption{Summary of CNN architecture used to estimate the neutral fraction from 21cm maps for the Simulated dataset. Same architecture is used for the mock dataset, except the input layer dimension is 40$\times$40.} 
    \centering 
    \begin{tabular}{l c} 
        \hline 
        Layer & Dimension \\ [0.5ex] 
        \hline 
        Input  (21cm map)& 200 $\times$ 200 $\times$ 1 \\ 
        2D Convolution & 200 $\times$ 200 $\times$ 8 \\
        2D Convolution & 200 $\times$ 200 $\times$ 8 \\
        Batch Normalization + Relu \\
        2D max-pooling & 100 $\times$ 100 $\times$ 8 \\
        2D convolution & 100 $\times$ 100 $\times$ 16 \\
        2D convolution & 100 $\times$ 100 $\times$ 16 \\
        Batch Normalization + Relu \\
        2D max-pooling & 50 $\times$ 50 $\times$ 16 \\
        2D convolution & 50 $\times$ 50 $\times$ 32\\
        2D convolution & 50 $\times$ 50 $\times$ 32 \\
        Batch Normalization + Relu \\
        2D max-pooling & 25 $\times$ 25 $\times$ 32 \\
        Flattening &  20000 \\
        Fully connected & 200 \\
        Batch Normalization + Relu \\
        Dropout \\
        Fully connected & 100 \\
        Batch Normalization + Relu \\
        Fully connected & 50 \\
        Batch Normalization + Relu \\
        Output (neutral fraction, x$_{\rm HI}$) & 1    \\    
        \hline 
    \end{tabular}
    \label{table:arch} 
\end{table}

\subsection{Training datasets}

\begin{figure}
 \centering
 \includegraphics[scale=0.4]{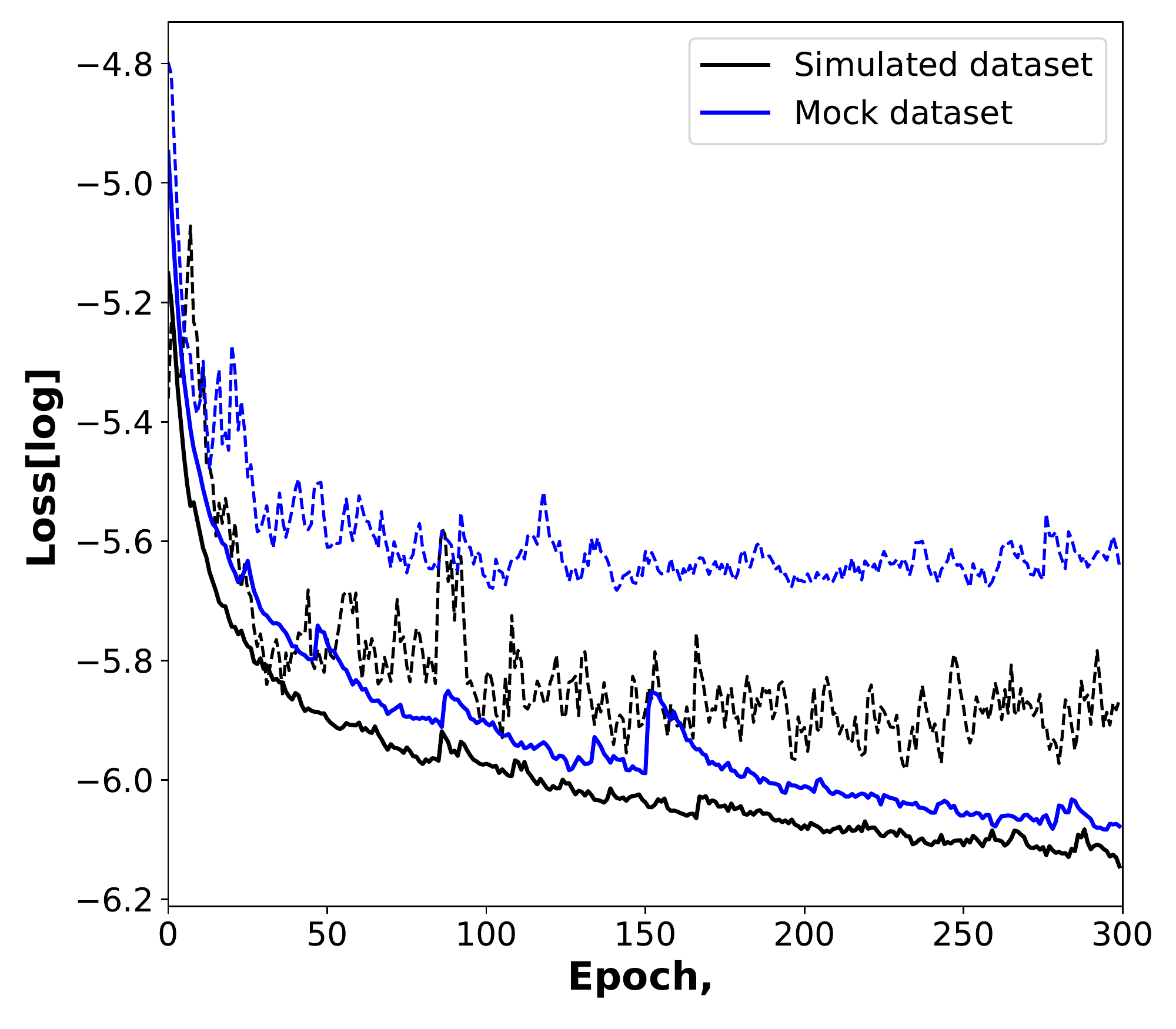}
 \caption{ The loss function (error rate) evolution over training epochs for training (solid) and testing (dashed) dataset without (black) and with (blue) instrumental effects from SKA. The network converges from the first 100 iterations for both training and testing with simulated dataset. Adding the instrumental effects from SKA increases the loss slightly, indicating slight reduction in overall accuracy.}
\label{fig:loss}
\end{figure}

\begin{figure*}
 \centering
 \hspace*{-1cm}\includegraphics[scale=0.6]{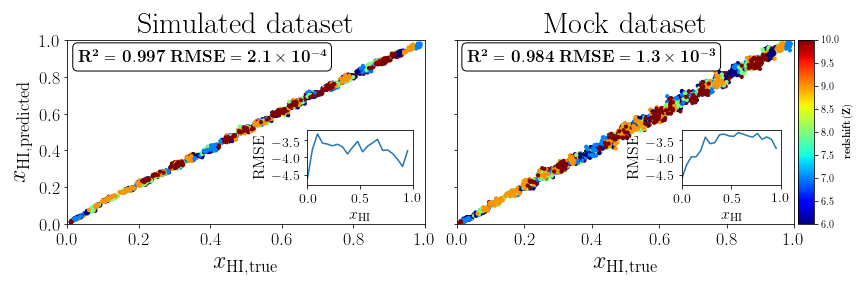}\hspace*{2cm}
 \caption{ Relationship between the true versus predicted neutral fraction values from the CNN for the simulated dataset (left) and mock dataset (right) using the whole validation dataset (circles) color-coded with redshifts which shows a weak dependence on the redshift. The insets show the RMSE evolution as a function of the neutral fraction, indicating moderate/strong evolution in the case of the simulated/mock dataset.  The total R$^{2}$ and RMSE values are quoted in each plot. For both datasets, the scatter is fairly small and the CNN predictions match the true values very well. Adding the SKA instrumental effects reduces the R$^{2}$ value only by 1 per cent.  }
\label{fig:acc}
\end{figure*}
\begin{figure*}
    \centering
    \includegraphics[scale=0.4]{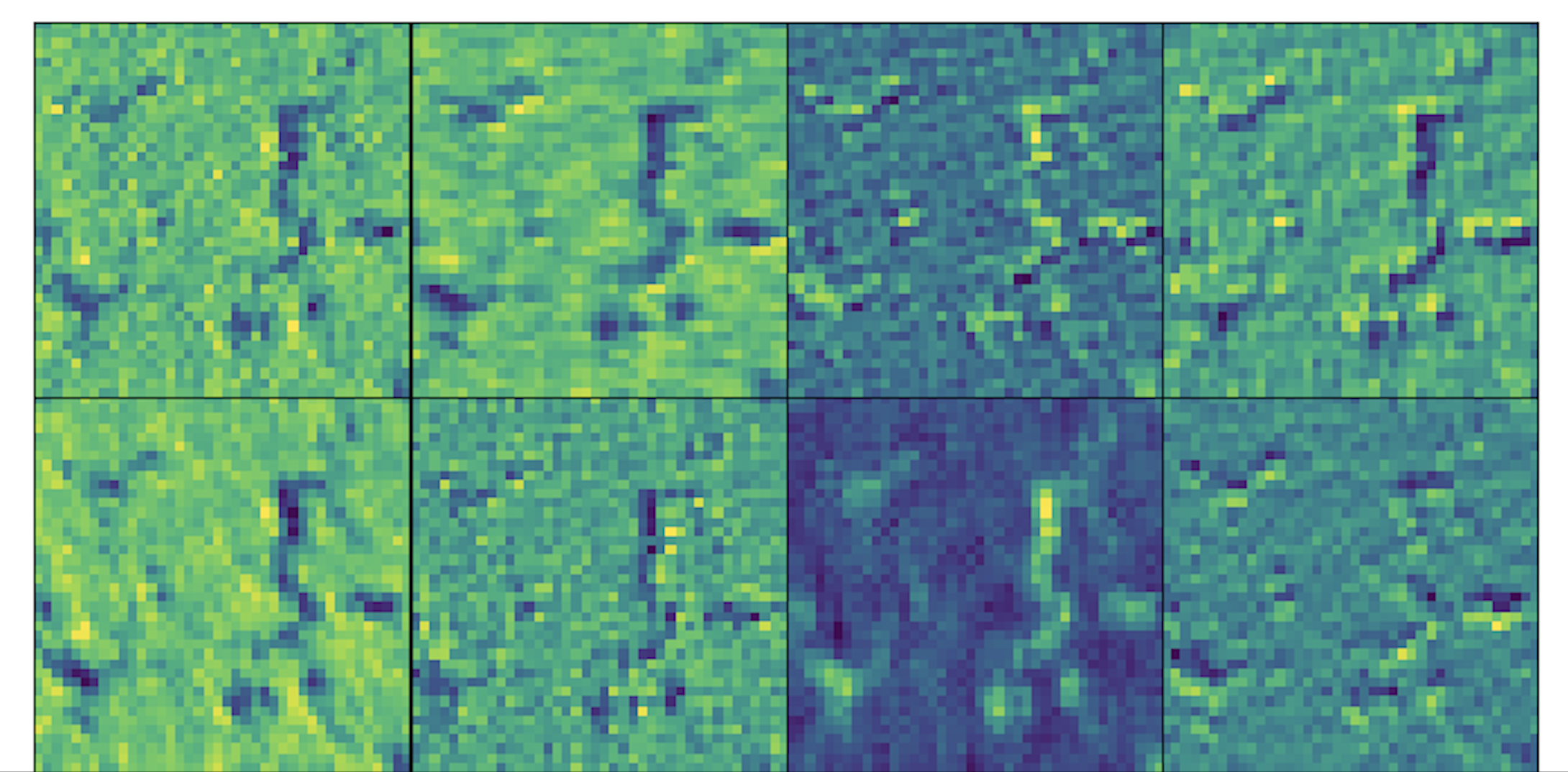}
    \caption{The response of convolving a randomly selected 21cm map with the trained set of 8 weights of the first convolutional layer before the application of the ReLU. The weights activate the input map differently, particularly, the neutral region. The ionized bubble activation appears to be approximately similar, albeit fainter with some weights. These variations are used in the network to estimate the neutral fraction out of the 21cm map.}
    \label{fig:filters}
\end{figure*}
Using {\sc SimFast21}, we run 1000 reionization simulations with a box size of 500 Mpc and number of cells N=200$^{3}$, which results in a resolution of 2.5 Mpc. Each realization is obtained from different realizations of the initial density field fluctuations through random change of the seed number, different set of astrophysical parameters, changing the photon escape fraction $f_{\rm esc}$=(0.01-1), $R_{\rm ion}$-M$_{\rm h}$ power dependence $C_{\rm ion}$=(0-1) and R$_{\rm ion}$ redshift evolution index $D_{\rm ion}$=(0-2), and different set of cosmological parameters, changing the matter density parameter $\Omega_{m}$=(0.2-0.4), the Hubble constant $H_{0}$=(60-80), and the  matter fluctuation amplitude $\sigma_{8}$=(0.7-0.9). The range considered for the astrophysical parameters is motivated from our previous MCMC estimates to match the simulation to several reionization observables~\citep{hassan2017epoch}, and those of the cosmology is inspired by the recent parameters estimates from the Planck Collaboration 2018. This ensures that our training sample contains very different set of 21cm maps that accounts self-consistently for the cosmic variance. For each simulation, we generate the 21cm brightness temperature box for redshifts=10,9,8,7 6, which are enough to form a decent size for training purposes. Overall, we extract  $
\sim  20,000$ 21cm maps in total and consider 80\%, 10\%, and 10\% out of this total for training, testing and validation, respectively. During the training, our dataset is divided into mini-batches of size 60. Note that these 21cm maps are labeled according to their neutral fraction, that is computed from the corresponding ionization map, instead of the globally averaged neutral fraction at each redshift. This indicates that our network aims to recover the local neutral fraction per slice by capturing the fluctuations along different directions.

\section{Results}\label{sec:results}

We train the network on the simulated dataset and then on the mock dataset with the presence of the SKA instrumental effects as described earlier. We first quantify network performance in terms of the evolution of the loss, alternatively called the error rate, over training iteration in Figure~\ref{fig:loss}. We show the loss for the simulated dataset with black and the mock dataset with blue color. Solid and dashed lines represent training and testing loss, respectively.  In both cases (Simulated or Mock), we find that the network converges very quickly from the first $\sim$ 100 training epochs where the training and testing losses both become approximately constant. The fluctuations around these losses are due to the random selection of mini-batch for training and testing at each iteration. When adding the instrumental effects from SKA as seen in the mock dataset, the loss value slightly increases for both training and testing. This slight loss change suggests a slight accuracy change, and hence our network design seems to be able to extract the neutral fraction information from large scale 21cm maps regardless the presence of SKA-like noise.

\begin{table}
    \caption{Root Mean Square Error (RMSE) and Coefficient of determination $(R^{2})$ for the simulated and mock validation datasets, as a function of redshift.} 
    \centering 
    \begin{tabular}{|c | c| c |c |c} 
        \hline 
        Redshift (z) & RMSE & $R^{2}$ & RMSE$_{\rm noise}$ & $R^{2}_{\rm noise}$\\
        \hline
        6 & 1.7$\times$10$^{-4}$ & 0.998 & 6.4$\times$10$^{-4}$ & 0.992\\
        7 & 1.2$\times$10$^{-4}$ & 0.997 & 4.5$\times$10$^{-4}$ & 0.995 \\
        8 & 1.9$\times$10$^{-4}$  & 0.998 & 1.1$\times$10$^{-3}$  & 0.986 \\
        9 & 2.3$\times$10$^{-4}$  & 0.997 & 6.8$\times$10$^{-4}$ & 0.991\\
        10& 2.4$\times$10$^{-4}$  & 0.997 & 3.7$\times$10$^{-3}$  & 0.955 \\ \hline
    \end{tabular}
    \label{table:res} 
\end{table}

We further assess the network performance using the coefficient of determination $R^{2}$ and the Root Mean Square Error (RMSE) as follows:
\begin{equation}
    R^{2} = 1 - \frac{\sum(y_{\rm predicted}-y_{\rm true})^{2}}{\sum(y_{\rm true} - \bar{y}_{\rm true})^{2}},
\end{equation}
\begin{equation}
    {\rm RMSE} = \sqrt{  \frac{1}{N} \sum(y_{\rm predicted}-y_{\rm true})^{2}   }\, ,
\end{equation}
where the summation runs through the whole validation dataset and the bar indicates the average. The $R^{2}$ quantifies the fraction by which the error variance is less than the true variance, whereas the RMSE measures the average squared error. In Figure~\ref{fig:acc}, we show the actual predictions for the neutral fraction from our designed CNN using the whole validation dataset. The color-code on the data points represents the redshifts as indicated on the right-hand side colorbar. We observe a very weak dependency of the neutral fraction on the redshift as there is no trend of colors-coded data points distribution visible (we note that we can have different neutral fractions at the same redshift since we are changing the simulations parameters). This is also seen from Table~\ref{table:res} where the R$^{2}$ and RMSE are approximately constant, particularly in the case for the simulated dataset. For the mock dataset, there seem to be a mild redshift evolution at high redshift as the RMSE drops by $\sim$ 1 order of magnitude between z=8-10. This is mainly driven by the increase of the thermal noise as a function of tedshift. In both panels in the case of the simulated and mock dataset, the scatter is fairly small which shows that our CNN is very successful in extracting the neutral fraction from 21-cm images with very high accuracy of $R^{2}$=99\% on the simulated dataset. When adding the instrumental effects from our assumed SKA design, the R$^{2}$ is only reduced by 1 per cent as shown in the right panel. The total RMSE  in estimating the neutral fraction, for the overall bins, is $2.1\times10^{-4}$ and $1.3\times10^{-3}$ for the simulated and mock datasets, respectively. Similar values are found for the RMSE. This shows that our designed CNN is a robust tool to constraining the neutral fraction, and hence the reionization history from future SKA observations. 

We next explore the neutral fraction recovery dependence on neutral fraction by computing the RMSE evolution for all neutral fraction bins as depicted in the insets in Figure~\ref{fig:acc}.
We find a strong evolution of the RMSE as a function of neutral fraction. The RMSE increases with increasing neutral fraction, indicating a harder parameter recovery. This is expected since the noise contaminates the signal more effectively at high neutral fraction values as seen in Figure~\ref{fig:sample}, where the ionized bubbles are very small. While no dependence on redshift is seen, it is known that the low neutral fraction IGM only exists at the end of reionization at z$\sim$ 7, 8. This shows that such a technique is very efficient to constraining the neutral fraction at low redshifts when the signal is low (due to low neutral fractions). 


We finally attempt to understand how CNN is able to create the link between the input (21cm-image) and the output (neutral fraction). We do so by looking at the response of the first convolutional layer at the final training step. As the images go deeper in the network (to the second and third layers), it becomes more difficult to visualize these images since some information will be lost and some transformation will occur, and hence we restrict the visualization to the first convolutional layer. In this first layer, we have a set of 8 weights, and we show the convolution of a random 21cm image to these weights in Figure~\ref{fig:filters}. We find that the activation of the ionized bubbles is relatively similar, although the bubbles edges are somewhat fainter with some weights. However, these trained weights do activate the neutral regions of the 21cm image very differently as shown by the red color, indicating that the network is using these variations to estimate the neutral fraction out of the input 21cm map.

\section{Conclusions}\label{sec:conclusion}

\noindent In this work, we have designed a convolutional neural network that is able to read out the neutral fraction from 21cm images, generated from our semi-numerical simulation ({\sc SimFast21}), with a very high accuracy  R$^{2} = 99$\%  and RMSE=$2.1\times10^{-4}$ (see Figure~\ref{fig:acc}). We have also considered the proposed SKA1-LOW array configuration to add realistic instrumental effects that account for the thermal noise, angular resolution and foreground contamination. We have shown that adding the SKA instrumental effects to the simulated 21cm images slightly delays the training process (see Figure~\ref{fig:loss}), but nevertheless the network is still able to extract the neutral fraction with a similarly high accuracy R$^{2}$=98\% and RMSE=$1.3\times10^{-3}$, which is only 1 per cent less than the accuracy without the presence of noise from the SKA (see Figure~\ref{fig:acc}). The accuracy depends weakly on redshift, but increases rapidly with decreasing neutral fraction. This is due to the fact that the instrumental noise increases towards high redshift where the Universe is highly neutral. The designed network activates the neutral regions in the 21cm images differently (see Figure~\ref{fig:filters}) which illustrates how the network creates the mapping between the input 21cm image and the output neutral fraction. 

Our results are limited to the approximation and assumptions used in the 21cm instrument simulations. A more refined recipe to account for the instrumental effects, particularly the foreground cleaning method, might alter our conclusion. The approximation implemented in the semi-numerical simulations place an additional limitation to the presented results.

The designed network shows the ability of machine learning to constrain the reionization history from 21cm-tomography with future EoR experiments. This approach is completely model independent as it relies on connecting directly images to ionization fractions.
The huge number of maps in the training dataset ($\sim$ 20,000) was very helpful to obtain such a high accuracy with/without the presence of instrumental effects from SKA. While we only focused on the SKA design, our analysis can be easily extended to include other experiments such as HERA and LOFAR, although the large noise and/or low resolution might create extra challenges. We leave for future work to present a detailed comparison between the ability of different 21cm arrays to constrain the reionization history. Our results also show that the foreground cleaning does not prevent the designed network to achieve very high accuracy. This is in contrast to previous results in~\citet{hassan2018identifying}, where the foreground cleaning can present a significant challenge to classify between galaxies and AGN reionization models, even for a highly sensitive 21cm array such as the SKA. In future work, we plan to check the network ability to constrain the reionization history from the 3D 21cm signal light cone, which contains more information than the static 21cm images used in this study.

\section*{Acknowledgements}
The authors acknowledge helpful discussions with M. Molaro, Arun Aniyan, Isaac Sihlangu, and Albert Baloyi. We particularly thank the anonymous referee for their comments which have improved the paper quality significantly. Simulations and analysis were performed at UWC's {\sc Pumbaa}, IDIA/{\sc Ilifu} cloud computing facilities, and NMSU's {\sc DISCOVERY} supercomputers. This work also used the Extreme Science and Engineering Discovery Environment (XSEDE), which is supported by National Science Foundation grant number ACI-1548562, and computational resources (Bridges) provided through the allocation AST190003P. MGS and TM acknowledge support from the South African Square Kilometre Array Project and National Research Foundation (Grant No. 84156).





\bsp	
\label{lastpage}
\end{document}